\documentclass[aps,prd,a4paper,showpacs,tightenlines,showkeys,preprintnumbers,amsmath,amssymb,12pt]{revtex4}
\usepackage{graphicx}
 \begin{document}
\title{Thermodynamical phases of a regular SAdS black hole}

\author{Anais Smailagic$^2$\email{anais@ts.infn.it}  Euro Spallucci$^3$\email{spallucci@ts.infn.it}}
\affiliation{$^2$ INFN, Sezione di Trieste, Italy
$^3$Dipartimento di Fisica, Universit\`a di Trieste and INFN, Sezione di Trieste, Italy}

\begin{abstract}
This paper studies the thermodynamical stability of \emph{regular}
BHs in  $AdS_5$ background. We investigate off-shell free energy of the system
as a function of temperature for different values of a ``coupling constant''
$\mathcal{L}=4\theta/l^2$, where the cosmological
constant is $\Lambda = -3/l^2  $ and $\sqrt\theta$ is a ``minimal length''.
The parameter $\mathcal{L}$ admits a \emph{critical} value, $\mathcal{L}_{inf}=0.2$, corresponding to
the appearance of an \emph{inflexion} point in the Hawking temperature.
In the weak-coupling regime $\mathcal{L} < \mathcal{L}_{inf} $, there are first order 
phase transitions at different temperatures. Unlike the Hawking-Page case, 
at  temperature $0\le T \le T_{min}$ the ground state is populated by ``cold''
near-extremal BHs instead of  a pure radiation.
On the other hand, for $\mathcal{L} > \mathcal{L}_{inf} $ only large, thermodynamically stable, 
BHs exist.

\end{abstract}


\maketitle
 
\section{Introduction}
 
The five dimensional Anti de Sitter geometry plays a central role
in recent developments in theoretical physics.  The $AdS_5/CFT$
duality \cite{Maldacena:1997re,Witten:1998zw,Witten:1998qj}
offers a powerful tool to tackle  non-perturbative features
of a variety of physical systems ranging from the quark-gluon plasma
\cite{Myers:2008fv}
to fluids \cite{Ambrosetti:2008mt} and super-conductors \cite{Hartnoll:2009sz}.
\\
The strong-coupling physics of a conformal field theory living
on the flat boundary of $AdS_5$ is mapped by duality to the weak-coupling
quantum string theory (quantum gravity) in the $AdS_5\times S_5$ bulk. 
This amazing spin-off of string theory connects $4D$ physics in 
flat spacetime to quantum gravity in $AdS_5$ in a beautiful realization
of the Holographic Principle 
\cite{Susskind:1994vu,Susskind:1998dq}.\\
In this framework the gravitational dual of the de-confinement phase 
transition turns out to be
the Hawking-Page transition between a cold gas of particles,
and a Schwarzschild AdS BH \cite{Hawking:1982dh}.
Thus, the study of thermodynamics of BHs in AdS
spacetime is instrumental to the deeper understanding of the 
above scenario.\\
 On the other hand, Schwarzschild AdS BH  suffers from the same pathologies as
 the corresponding solution without cosmological constant. 
For example,  Hawking temperature is divergent as the horizon is shrinks to zero, 
and there is a curvature singularity at the origin of the coordinate system.
\\
Recently, we have found pathology-free, regular BH solutions of 
Einstein equations, that led to new predictions about the terminal phase of quantum evaporation 
\cite{Nicolini06,Ansoldi:2006vg,Spallucci:2008ez,Nicolini:2008aj,Smailagic:2010nv,Nicolini:2009gw}.
In a recent paper \cite{Mann:2011mm} regular BH creation in a deSitter
background geometry was studied and it was found that these objects
 would have  been plentifully produced during
inflationary times, for Planckian values of the cosmological constant.\\
Motivated by the promising results quoted above, we would like to analyze
thermodynamical behavior of neutral, non-spinning, \emph{regular} BH in $AdS_5$ background
with $\Lambda =-3/l^2 $.\\
The paper is organized as follows: in Sect.(2) we review the $5D$
Einstein equations with a Gaussian source and a negative cosmological
constant. Regular Schwarzschild $AdS_5$ BH is introduced
and discussed from a geometrical point of view. In Sect.(3) we pave the
way to the study of thermodynamics of the solution presented in Sect(2).
and compute the Hawking temperature, the BH entropy and Off-Shell Free Energy.
In Sect.(4) we study the phase transitions in our model.  Two
regions of the ``~coupling constant~'' $\mathcal{L}=4\theta/l^2$ are described. A rich phase structure 
emerges in the weak-coupling regime $\mathcal{L}<\mathcal{L}_{inf}=0.2  $,   
with important differences with respect to the
Hawking-Page scenario  in the near-extremal region. The resulting picture closely resembles first order phase
transitions in a finite temperature quantum field theory\cite{Lombardo:2000rs,ogilvie}. On the other hand
in the ``string-coupling'' regime $\mathcal{L}>\mathcal{L}_{inf}=0.2 $ there is always a single BH at any
temeperature. 
Finally, in Sect(5) we summarize the results obtained and stress the novel features.

\section{Regular AdS BH.}

In this Section a brief review of the ideas underlying regular
BH  solutions of Einstein equations is given. We  work in
$5D$ spacetime being the proper dimension for the $CFT/AdS$ Correspondence.\\
 The line element is a static, spherically
symmetric solution of the $5D$ Einstein equations, with a negative cosmological
constant $\Lambda\equiv -3/l^2$(AdS). For the positive cosmological constant
case, see \cite{Mann:2011mm}. The energy momentum tensor leading to
the regular metric is of the anisotropic fluid form
\cite{Nicolini06}, with the Gaussian matter source. This type of matter 
distribution emulates non-commutativity of space-time through a parameter $\theta$
corresponding to the area of an elementary cell. Its components are given by
 
\begin{eqnarray}
&& T^0_0=T^r_r=-\rho\left(\, r\,\right)\ ,\quad
\rho\left(\, r\,\right)\equiv \frac{M}{\left(\, 4\pi\theta\,\right)^2}
\exp\left(\, -r^2/4\theta\,\right)\ ,\\
&& T^m_m=-\rho -\frac{r}{3}\partial_r \rho
\end{eqnarray}
 where $M$ is a total mass-energy of the source and
$\sqrt\theta$ is a natural UV cut-off curing short distance
infinities. These ideas, based on UV divergence-free field theory models described in
\cite{Smailagic:2003yb,Smailagic:2003rp,Smailagic:2004yy,Spallucci:2006zj}
and applied to a quantum
gravity model, allow interpretation of parameter $\theta$ in different terms: 
  string induced non-commutative geometry \cite{Seiberg:1999vs},
  zero-point length \cite{Fontanini:2005ik,Spallucci:2005bm},
  etc.,  depending on the assumed underlying fundamental quantum gravity theory.\\
The advantage of our approach is that,
contrary to the star-product approach description, there is no need to modify
  the metric part of Einstein equations. Rather,  ''quantum'' effects
  are incorporated through the energy-momentum tensor.
  Since matter induces metric, solving equations will modify geometry
  appropriately. Thus, starting from Einstein equations
 
\begin{equation}
R^M_N -\frac{1}{2}\left(\, R-2\Lambda\, \right)\, \delta^M_N=8\pi\, G_5
T^M_N\ , \quad \Lambda\equiv -\frac{3}{l^2}\label{e5}
\end{equation}
 
with, $G_5=1/M_\ast^3$  the $5D$ gravitational coupling constant.
The solution is the line element
 
\begin{eqnarray}
&& ds^2 = - \left[\, 1-\frac{2G_5M\left(\, r\,\right)}{r^2}
+\frac{r^2}{l^2}\,\right]\, dt^2
+ \left[\, 1-\frac{2G_5M\left(\, r\,\right)}{r^2}
+\frac{r^2}{l^2}\,\right]^{-1} dr^2\nonumber\\
&&\quad\quad\quad + r^2\, d\Omega_3
\label{sol}\ ,\\
&& M\left(\, r\,\right)\equiv  M\,\gamma\left(\,2\ ,
\frac{r^2}{4\theta}\,\right)
\end{eqnarray}
 
where, $d\Omega_3$ is 3-sphere line element and $\gamma\left(\,2\ ,
r^2/4\theta\,\right)$ is the incomplete gamma function defined as
 
\begin{equation}
\gamma\left(n\ , r^2/4\theta\, \right)\equiv
\int_0^{r^2/4\theta} dt\, t^{n-1} e^{-t}
\end{equation}
 
 The short-distance regularity of Eq.(\ref{sol}) can be 
 inferred from the behavior of the incomplete gamma at short distance
 
 \begin{equation}
 \gamma\left(n\ , r^2/4\theta\, \right)
 \approx \frac{1}{n}\left(\, \frac{r^2}{4\theta}\,\right)^n
 \end{equation}
 
 Physically, this means that spacetime fuzziness smears the 
 curvature singularity creating either Minkowski or (A)dS 
 metric. Different possibilities are determined by the
 interplay between UV and IR physics characterized by $\theta$
 and $\Lambda$ parameters respectively. Giving opposite contributions
 to the short distance energy density leads to three different
 situations: fine tuning of parameters leads to Minkowski metric; 
 prevailing $\theta$ gives deSitter, while prevailing $\Lambda$
 leads to AdS. \\
 The short-distance behavior allows an intriguing interpretation in terms
of the effective gravitational coupling as follows: let us introduce 
  $G_5^{eff.}$ as
 
\begin{equation}
G_5\longrightarrow G_5^{eff.}\left(\, r\,\right)\equiv
G_5\gamma\left(\, 2\ ,r^2/4\theta\,\right)
\end{equation}
 
The effect  of the smeared coordinates can be interpreted as the vanishing 
of the  ``\textit{running}''  gravitational constant as $r\to 0$ . Thus, our model  
can be viewed as an effective \textit{asymptotically free} gravitational model
as shorter distances are probed.
We remark that the weakening of gravity  leads \textit{only} to
the absence of curvature singularities but does not prevent formation
of BHs. Even in the  case of the extremal BHs, the horizon radius
$r_0$ is still few $\sqrt\theta$ units and
$G_5^{eff}\left(\, r_0\,\right)$ is  big enough to create an event
horizon.  \\
One of the novel features of the our regular solution is the existence of
\emph{multiple} horizons, unlike ordinary Schwarzschild solution.
Unfortunately, due to the complicated metric function, horizon equation can
not be solved explicitly. Nevertheless, the existence of zeros in the
$g_{rr}^{-1}$ component of the metric can be determined from
the graph of the function $M=M\left(\, r_H\,\right) $ defined
from $g_{rr}^{-1}\left(\, r_H\,\right)=0$. Thus, one gets
 
\begin{equation}
M=\frac{r^2_H}{2G_5\gamma\left(\, 2\ ,r^2_H/4\theta\,\right)}
\left(\, 1+\frac{r^2_H}{l^2}\,\right)
\label{adm}
\end{equation}
The above equation is to be interpreted as follows:
$M$ plotted as function of $r_H$ 
describes the existence of horizons in the sense that for an assigned value of 
$M$ the horizons are given by
intersections of $M=const$ with the graph
$M\left(\, r_H\,\right)$.  In the next section
$M(r_H)$ will be given a  thermodynamical interpretation  as the
\textit{Internal Energy} of the system depending on the
(outer, Killing) horizon $r_H=r_+$.
\begin{figure}[ht!]
 \begin{center}
\includegraphics[width=15cm,angle=0]{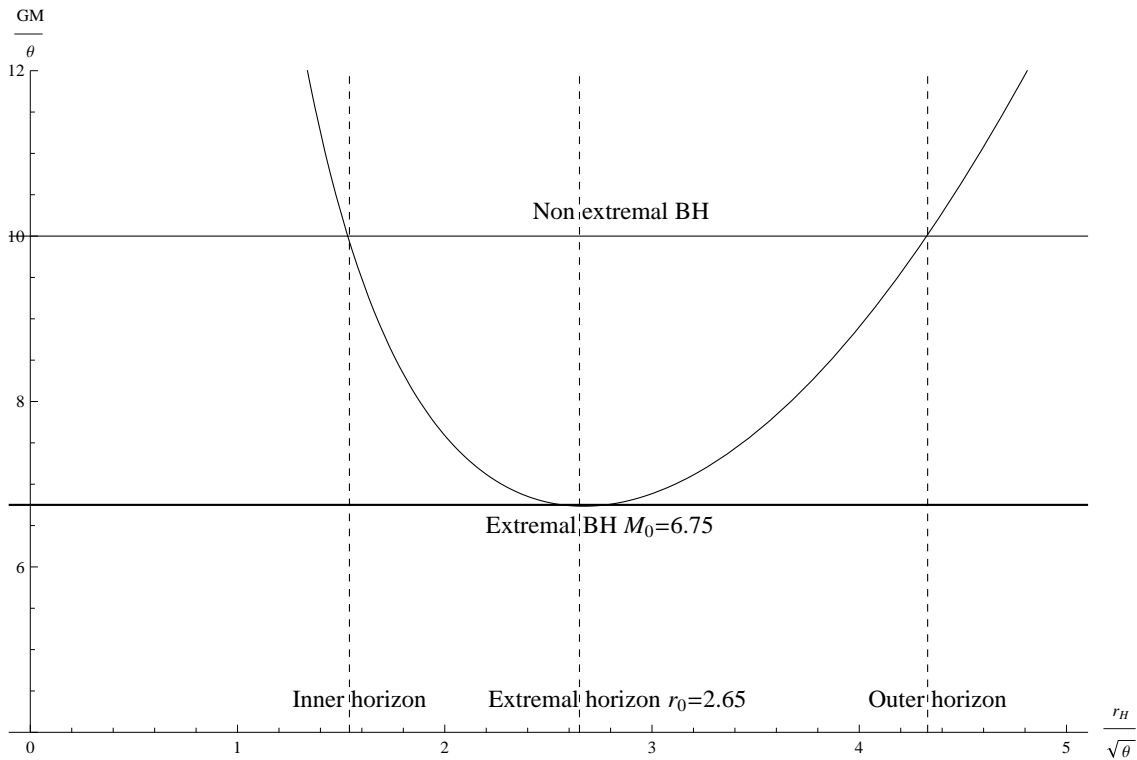}
\caption{\label{admmass} Plot of the function $MG_5/\theta$ vs $r_H/\sqrt\theta$, with $3\theta/l^2=0.002$.
 $r_0=2.65\sqrt\theta $ is the radius of the \emph{extremal} BH with mass
$M_0 =6.75\,\theta/G_5$.
For any $M>M_0$ there exist an inner (Cauchy) and an outer (Killing) horizon $r_H=r_\pm$ as shown
by the horizontal line $MG_5/\theta=10$. }
\end{center}
\end{figure}
 
The graph  of equation (\ref{adm}) is shown in Fig.(\ref{admmass})
The existence of a minimum indicates an \textit{extremal} BH of
radius $r_0$ and mass $M_0$ for a given $\Lambda$. 
Therefore, regular BHs admit an \emph{extremal} configuration, contrary to standard Schwarzschild AdSBH. 
For $M> M_0$ there are \textit{non-degenerate}
BHs with distinct inner (Cauchy) horizon and outer (Killing) horizon.
For $M< M_0$ there is no horizon and the solution (\ref{sol}) represents
a maximally localized mass in an AdS background.\\

The existence of a stable, minimal size,  BH has interesting
consequences which have been discussed in \cite{Spallucci:2011rn}.
If we assume the free parameter $M$ in the metric to be the mass of an ``elementary'' object,
then the solution (\ref{sol}) can be considered as a family containing
\textit{particle-like}, light
objects for $M<M_0$ and \textit{black}, massive object for $M>M_0$.
Extremal mass
$M=M_0$ is the transition energy between particles and BHs.
The described behavior of regular BH goes hand-in-hand with recent ideas
posing a limit of any physical high energy probe
\cite{Dvali:2010bf,Dvali:2010ue}.
The following scenario is in place: the size of a
particle is characterized by its
Compton wavelength and is a \emph{ decreasing} function of  mass. Once the
extremal value $M_0$ has been reached, any increase in $M$ will lead to the
\emph{increase}
of the radius of the corresponding BH. Thus, $\lambda_C=1/M_0$
is the minimum wavelength for any physical high-energy probe. From
this perspective, $r_0$ plays the role of the \textit{effective} Planck
length and there is no way to explore shorter distances \cite{Spallucci:2011rn,Spallucci:2012xi,Nicolini:2012fy}.

\section{Thermodynamics}
 \begin{figure}[ht!]
\begin{center}
\includegraphics[width=15cm,angle=0]{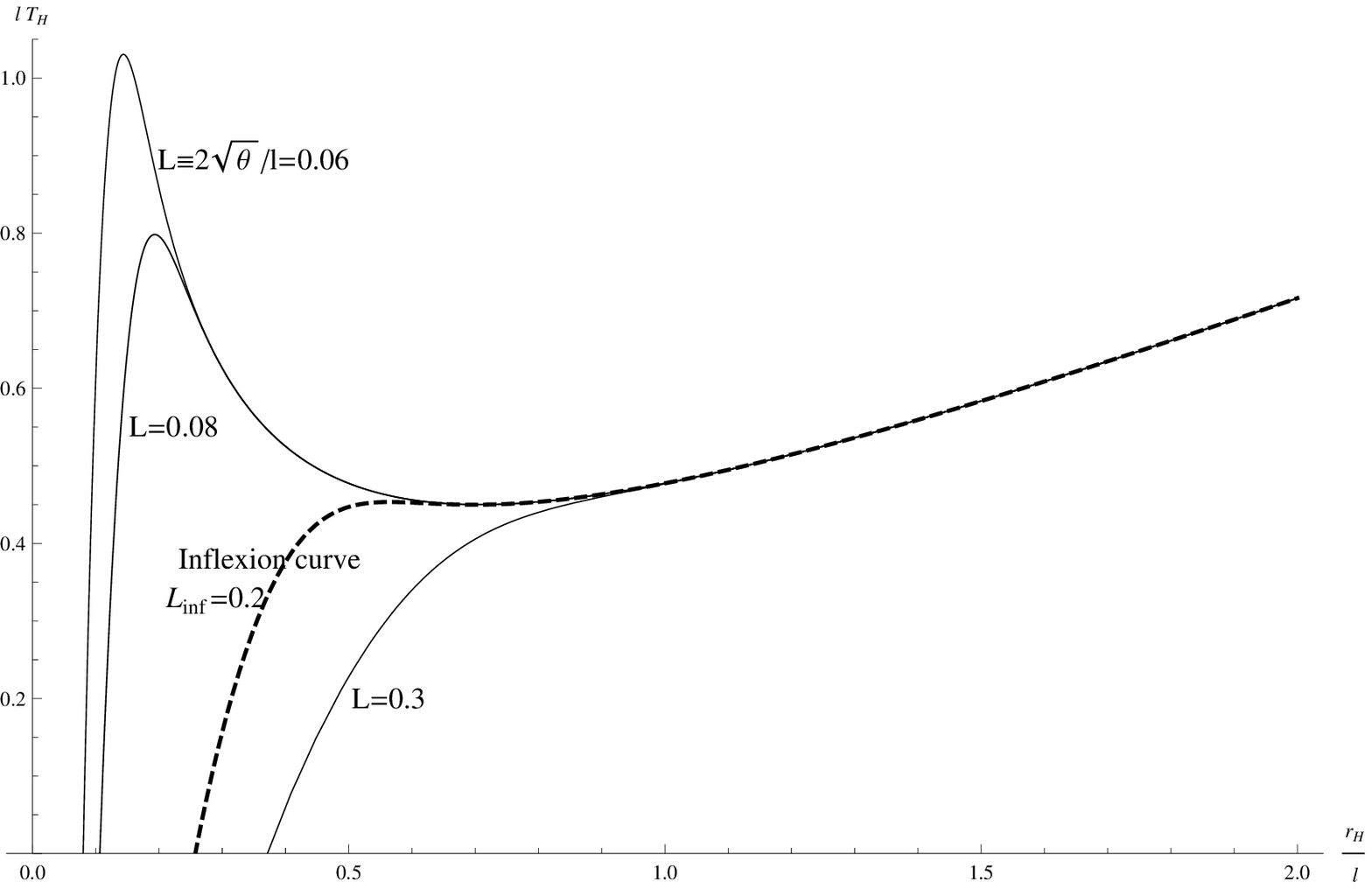}
\caption{ $ l T_H$ vs $r_+/l $ for different values
of the  parameter  $\mathcal{L}\equiv 4\theta/l^2$.\\
For $0 < \mathcal{L} < L_{inf}=0.22 $ the temperature has a local maximum and minimum. At the
\emph{critical} value $ \mathcal{L}_{inf}=0.22 $ the two extrema merge into an \emph{inflexion} point (dashed curve).
Above the critical value, $T_H$ is a monotonically increasing function of $r_+$.
}\label{tmp4}
\end{center}
\end{figure}

Thermodynamical description   of regular BH (\ref{sol})
will follows the same steps as in \cite{Nicolini06}.
The Hawking Temperature of the AdS BH is given by
 
\begin{equation}
T_H=\frac{1}{2\pi\,r_+}
\left[\frac{\gamma\left(\, 3\ ,r^2_+/4\theta\,\right)}
{\gamma\left(\, 2\ ,r^2_+/4\theta\,\right)}\left(1+\frac{r_+^2}{l^2}\right)
-1\right]
\label{tads}
\end{equation}

The metric (\ref{sol}) admits an extremal BH configuration even without being charged or rotating. This
is a general property of regular solutions sourced by a finite width $\sqrt\theta$, Gaussian energy/momentum distribution.
One can see from the relation

\begin{equation}
 4\pi T_H = \frac{2G_5}{r_+^2} \, \Gamma\left(\, 2\ , r_+^2/4\theta\,\right)\frac{dM}{dr_+}
\end{equation}

that $T_H$ vanishes at the radius $r_0$ which minimizes $M\left(\, r_+\,\right)$.\\
Furthermore, equation (\ref{tads}) displays   $T_H$ dependence  on two length scales, 
i.e. the short-distance cut-off
$\sqrt\theta$ and the AdS curvature radius $l$. \\
By letting $\theta \to 0$ and using fundamental properties of the Euler Gamma function,
 we recover the  temperature of a $5D$ SAdS BH

\begin{equation}
T_H=\frac{1}{\pi\,r_+}
\left[\left(1+\frac{r_+^2}{l^2}\right)
-\frac{1}{2}\right]
\label{tsads0}
\end{equation}

If we let the AdS radius to infinity, i.e. vanishing of the cosmological constant, $\Lambda\to 0 $, we get
the temperature of a regular, $5D$, SAdS BH

\begin{equation}
T_H=\frac{1}{2\pi\,r_+}
\left[\frac{\gamma\left(\, 3\ ,r^2_+/4\theta\,\right)}
{\gamma\left(\, 2\ ,r^2_+/4\theta\,\right)}
-1\right]
\label{trsads}
\end{equation}

 In order  to study the behavior of the temperature in intermediate situations, i.e.
for a finite, non-vanishing, 
ratio between the two length scales, we rescale equation (\ref{tads}) as follows
\begin{equation}
lT_H=\frac{1}{2\pi\,x}
\left[\frac{\gamma\left(\, 3\ ,x^2/\mathcal{L}^2\,\right)}
{\gamma\left(\, 2\ ,x^2/\mathcal{L}^2\,\right)}\left(1+ x^2\right)
-1\right]\ ,\qquad \mathcal{L}^2\equiv \frac{4\theta}{l^2}\ ,\qquad x\equiv\frac{r_+}{l}
\end{equation}

If we assign $\mathcal{L}$ the meaning
of ``\emph{coupling constant}'', we can introduce a ``\emph{critical coupling}'' $\mathcal{L}_{inf}=0.2$ 
for which $T_H$ has an \emph{inflexion point}. Thus, one can define two \emph{different regimes}:
\begin{itemize}
\item ``\emph{weak-coupling}'' 
corresponding to $\mathcal{L}< \mathcal{L}_{inf}$;
\item ``\emph{strong-coupling}'' for $\mathcal{L}> \mathcal{L}_{inf}$.
\end{itemize}
We shall show in the next sections
that these two regimes correspond to different behavior of the BH.\\
In order to investigate the thermodynamical equilibrium of this gravitational
system, we are going to study the behavior of the off-shell (~Helmoltz~) free energy.
To interpret the results we shall adhere to the description of phase transitions used in
\textit{finite temperature field theory} \cite{Lombardo:2000rs,ogilvie}. 
In our case, free energy plays the role of the effective potential and $r_+$ is the order parameter.
We investigate \textit{off-shell} free energy, $F^{off}$,
 since it describes the non-equilibrium
dynamics of a system in a thermal bath of temperature $T$, which is a free
parameter and should not to be confused with Hawking temperature $T_H$.  In this way, one can study the
evolution of the system towards a stable equilibrium configuration, $T=T_H$, and eventual
phase transitions.  $F^{off}$ is defined as

\begin{equation}
F^{off}\equiv M\left(\, r_+\,\right) -S_H \, T
\label{foff}
\end{equation}
 
 where, $S_H$ is the BH entropy determined from the First Law as
 
\begin{equation}
  dM\left(\, r_+\,\right)\equiv T_H dS_H
  \longrightarrow S_H=\frac{2\pi}{G_5}\int_{r_0}^{r_+} dr
  \frac{r^2}{\gamma\left(\, 2\ ,r^2/4\theta\,\right) }\label{entropia}
  \end{equation}

Integration has to start from the extremal radius $r_0$, rather than from zero.
This choice automatically guarantees 
vanishing thermodynamical entropy at absolute zero, as it is required by the 
Third Law.
Furthermore, we are not \textit {a priori} assuming
the Area Law, rather we shall derive the relation
between Entropy and Area \textit{from} the First Law.
Integrating by parts in (\ref{entropia}) one obtains
 
\begin{eqnarray}
&& S_H =\frac{1}{3\pi}\left[\,\frac{2\pi^2\, r_+^3}{G_5^{eff.}\left(r_+\,\right) }-
\frac{2\pi^2\, r^3_0}{G_5^{eff.}\left(\, r_0\,\right)}\,\right]
+\Delta S_H\nonumber\\
&& \Delta S_H =\frac{\pi}{12\theta^2\, G_5}\int_{r_0}^{r_+}dr r^6
\frac{e^{\left(\,-r^2/4\theta\,\right)}}
{\gamma^2\left(\, 2\ ,r^2/4\theta\,\right)}
\label{entropy}
\end{eqnarray}

One recognizes that the first line of (\ref{entropy}) is the   
Area Law for regular, 5D BHs, while 
the second line gives exponentially small corrections  
\cite{Spallucci:2008ez,Nicolini:2008aj}. 
Standard Area Law is obtained only in the limit 
$\sqrt\theta\to 0\ ,G_5^{eff.}\to G_5 $.
 By inserting $S_H$ from Eq.(\ref{entropia}) in Eq.(\ref{foff}), we obtain
\begin{equation} 
 F^{off}= \frac{r_+^2}{2G_5\gamma\left(\, 2\ ; r^2_+/4\theta\,\right) }
 \left(\, 1+ \frac{r_+^2}{l^2}\,\right) 
 -\frac{2\pi\, T}{G_5}\int_{r_0}^{r_+} dr
  \frac{r^2}{\gamma\left(\, 2\ ,r^2/4\theta\,\right) }
 \label{foff2}
 \end{equation}
 
 It is not possible to  integrate analytically the entropy expression,
 and to write a closed form for (\ref{foff2}). Nonetheless, it is possible
 to plot (\ref{foff2}) and the resulting graphs are shown in next sections.


\section{BH Phases }
 \subsection{Weak-coupling phase}
\begin{figure}[ht!]
\begin{center}
\includegraphics[width=15cm,angle=0]{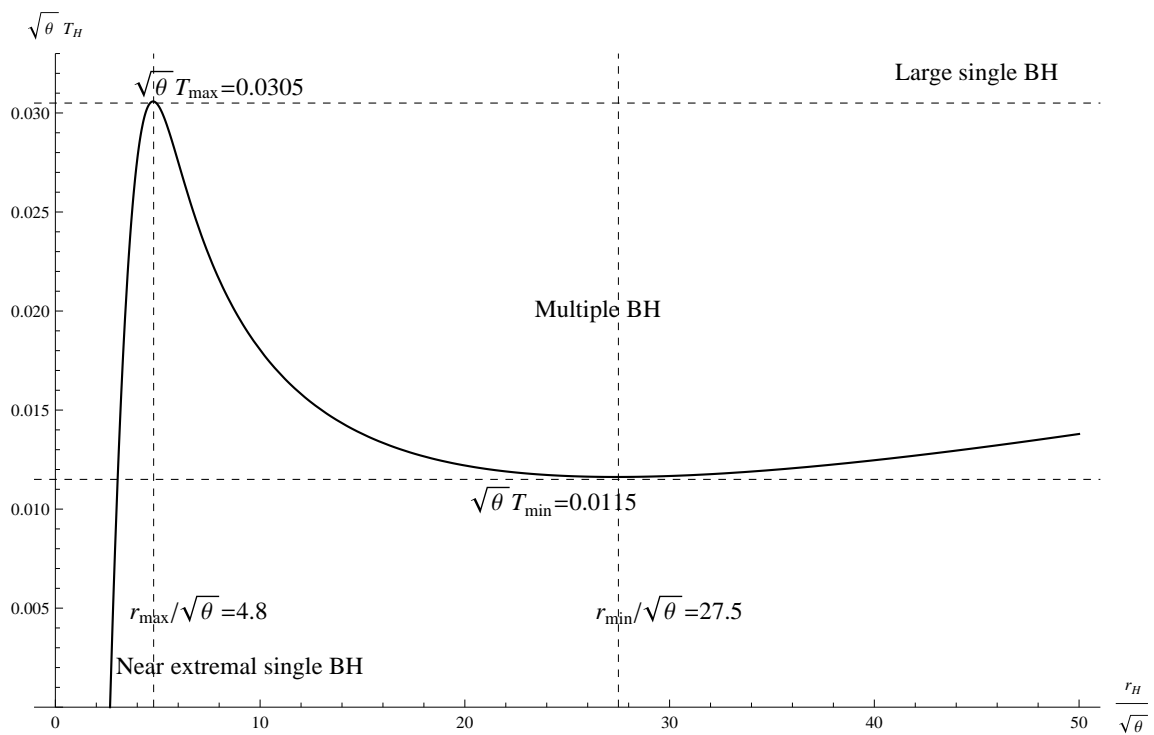}
\caption{\label{temp}Plot of the rescaled Hawking Temperature $\sqrt\theta T_H$ vs. $r_+/\sqrt{\theta}$. 
The curve corresponds to the value $\mathcal{L}^2=0.002$. The temperature vanishes for the extremal 
configuration $r_H=r_0$, then reaches a local maximum in $r_{max}=4.8\sqrt\theta $, further decreases 
to a local minimum in $r_{min}=27.5\,\sqrt\theta $ and finally raises linearly.
For $0\le T < T_{min.}$ we have small, stable, \emph{near-extremal}, BHs. 
In the range $T_{min.}< T < T_{max}$, 
 we find multiple (triplet) BHs:
 small, stable, near-extremal BH; intermediate radius , unstable, BH; large, stable, BH.  
Finally, for $ T >T_{max}$, there is a single, large, stable BH. }
\end{center}
\end{figure}

\begin{figure}[h!]
\begin{center}
\includegraphics[width=15cm,angle=0]{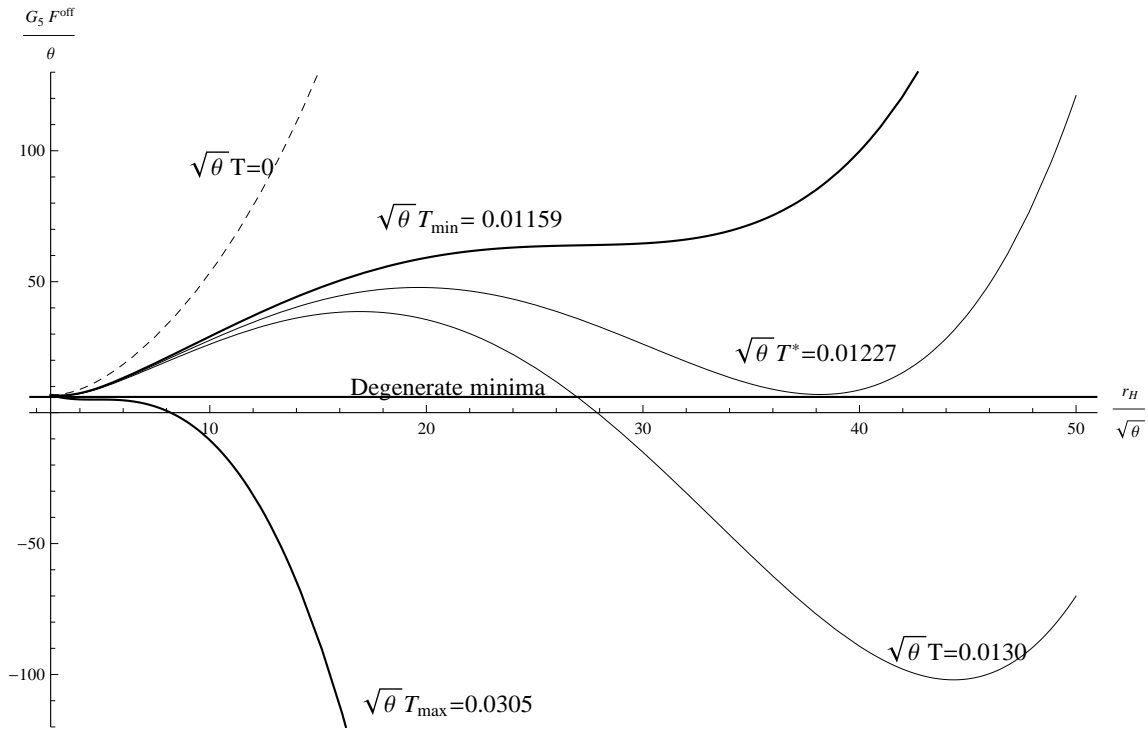}
\caption{Graph of the rescaled Off-shell Free energy, for $\mathcal{L}^2= 0.002$, for different
values of the temperature $T\sqrt\theta$. At the  temperature
$T^\ast\sqrt\theta =0.01227$ the two local minima are degenerate.  Two inflexion points correspond to
the maximum and minimum of the Hawking temperature.}\label{free1}
\end{center}
\end{figure}

\begin{figure}[h!]
\begin{center}
\includegraphics[width=15cm,angle=0]{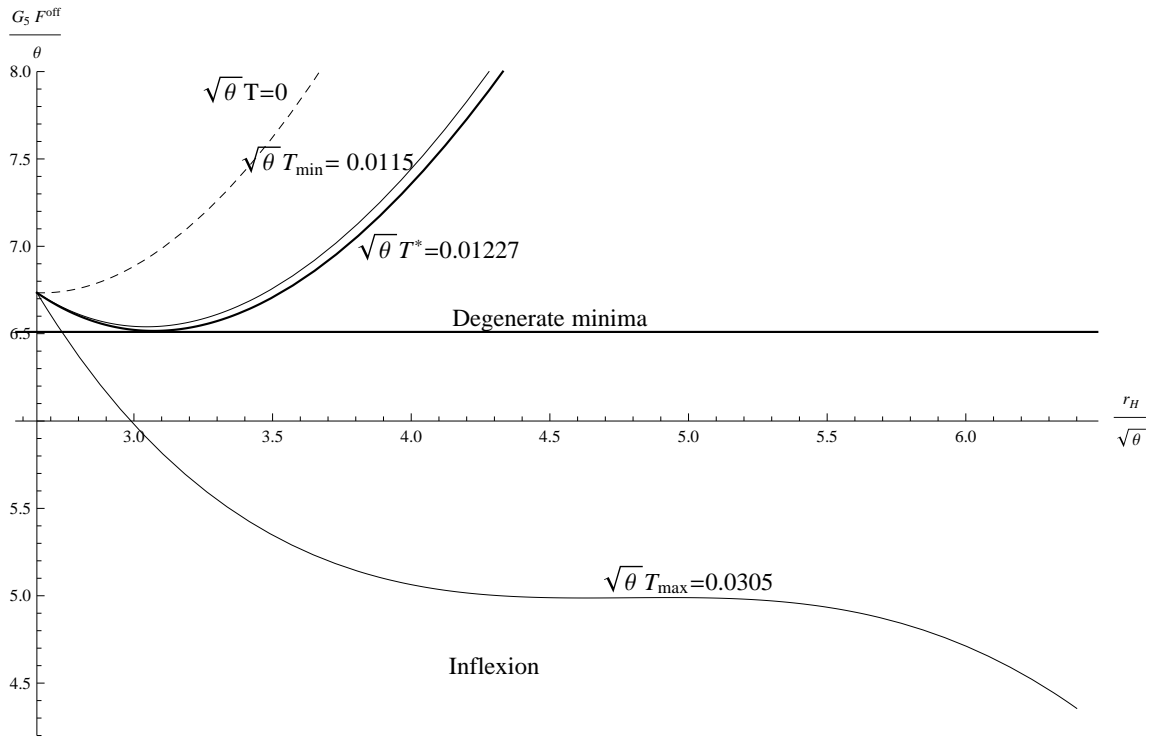}
\caption{This is a zoom of the previous figure in the 
near-extremal region. For  $0\le \sqrt\theta\, T  < 0.0305 $ a new local minimum appears describing a small,
near extremal, BH. For $ \sqrt\theta\, T =T_{max.} = 0.0305 $ the local minimum and maximum of
$F^{off}$ merge into an inflexion point. 
For $ \sqrt\theta\, T  > 0.0305 $ there is  only a large BH  representing a stable
ground state of the system.
}\label{zoom}
\end{center}
\end{figure}

Fig.(\ref{temp})  indicates the existence of multiple BHs with the same temperature, as well as
extremal BH. As the extremal configuration is approached, a sharp departure
with respect to the standard AdS BH occurs.
The behavior of the temperature graph can be summarized as follows:
extremal configuration  corresponds to zero temperature
and can be considered true ground state of the system with minimized internal
energy  $M\left(\,r_+\,\right)$. In the near extremal region $T_H$ increases from zero
to a local maximum $T_{max}$ at $r_+ = r_{max}$. As $r_+$  increases further $T_H$
drops to a local minimum  $T_{min}$ corresponding to $r_+ = r_{min}$ and then
starts increasing linearly following the SAdS pattern.\\
The behavior of $F^{off}$  is shown in Figure (\ref{free1}).
Comparison between the graph of (\ref{foff2}) and the same expression for $\theta=0$
(Hawking-Page AdS),   shows  a novel feature of our regular solution. 
  It develops a \textit{new}, local, near-extremal minimum, for any 
  $T>0$ as seen in Figure(\ref{zoom}). This behavior is a
consequence of the smearing of the matter source resulting in the
existence of the extremal horizon.\\

The extrema of free energy indicate existence of both \textit{multiple} and 
\emph{single} regular BHs for different values of the temperature. 
The alternation of single/multiple states is the signature of a first order phase transition,
as in finite temperature quantum field theory, to which we refer.\\ 
It turns out that single/multiple BH transitions occur at the inflection
points of free energy (~extremal points of $T_H$~).
 In order to grasp better what is going on in the near extremal region, we zoom that part of
 Figure(\ref{free1}) in Figure(\ref{zoom}).
Thus,  the following scenario is in place:\\
\begin{enumerate} 
\item
$T=0$. BH is in the \textit{frozen single state}. 
The only ground state is the extremal configuration with $r_+=r_-=r_0$.
\item $0\le T \le T_{min}$. BH is in the \textit{cold single} state. The
stable equilibrium configurations are small, \textit{near-extremal},
BHs of radius $r_0< r_+ < r_{min}$. This situation is completely novel with
respect to  the Hawking-Page scenario, which is characterized by a gas of cold particles
in this phase. BHs can not be formed below some minimal temperature. 
On the contrary, in our case, BHs are always present, as soon as 
$T\geq 0$. This is due to the effect of the minimal length $\sqrt{\theta}$. 
\item
$T=T_{min}$ corresponds to the onset of a \textit{spinodal decomposition} in
quantum field theory.  
An inflexion point appears in $F^{off}$ at $r_+=r_{min}$ (see Figure(\ref{free1})). 
\item
  $T_{min} < T < T^\ast$. New local minimum develops and the system
splits into two  co-existing states. The small
near-extremal BH is energetically favored.  
\item
$T = T^\ast$ the two minima become degenerate
and the system is in a \textit{mixed state}. Both BHs have the same free energy.
\item $T^\ast < T < T_{max}$ large BHs become stable, while near-extremal 
 BHs are only locally stable.
\item $T=T_{max}$ The near-extremal minimum merges with the local maximum.
There is a  new  transition from multiple to a single BH state.  
\item $T>T_{max}$  there is \textit{high temperature, single, stable} BH.
\end{enumerate}
The  above scenario describes  \textit{first order} 
phase transitions from single to multiple BHs at $T=T_{min}$ and $T=T_{max}$.

\subsection{Strong-coupling phase}

\begin{figure}[ht!]
\begin{center}
\includegraphics[width=15cm,angle=0]{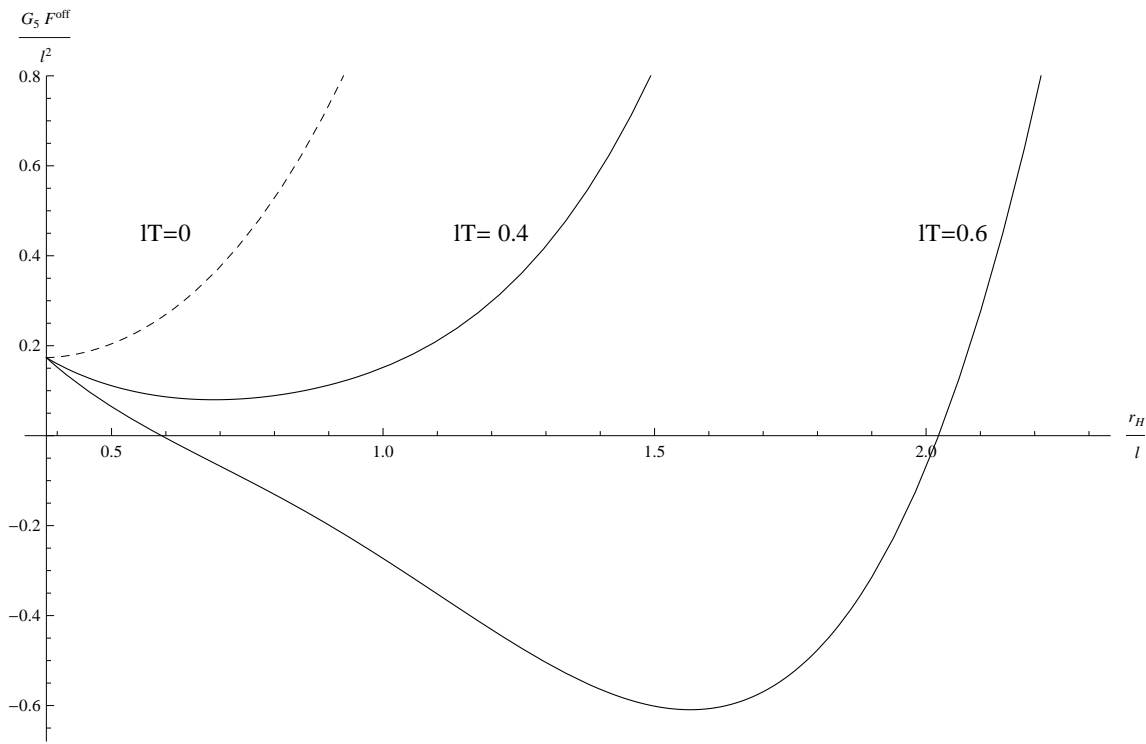}
\caption{Free energy $G_5 F^{off}/l^2$ vs $r_+/l$. For $L\ge 0.22$
there exist only one minimum for any $T$. }
\label{tmp5}
\end{center}
\end{figure}

One realizes that the above description of first order phase 
transitions is only correct in the weak-coupling regime.
 In fact, the two parameters of the theory have 
``opposite'' effects i.e. $\theta$ dominates short-range and lowers $T_{max}$, while $3/l^2$ 
dominates long-range region of $r_H$ and raises $T_{min}$.\\
It is reasonable to expect that, at the certain point, these two opposing 
effects will meet creating an \textit{inflexion} point of temperature.
The confirmation of our conjecture is shown in Fig.(\ref{tmp4}). 
This effect, in terms of free energy, means that, beyond the inflexion point ${\mathcal{L}}_{inf}=0.2 $, 
there is only one minimum at \textit{any} value of parameter $T$ and thus, only one BH state.
Changing $T$ only changes \textit {position} of the \textit{single} minimum. 
The graph of free-energy in the strong-coupling regime is given in 
Fig.(\ref{tmp5})

\section{Conclusions}
 
 We have investigated thermodynamical stability of \textit{regular} BH in AdS 
 background. The thermal bath of particles is described by the temperature 
 parameter T and is not \textit{a priori} in thermal equilibrium with BH. 
 We studied the behavior of off-shell free energy in order to study evolution 
 towards equilibrium. In this picture, thermal equilibrium states are described 
 by minima of $F^{off}$ ( in this paper we have not considered 
 quantum instabilities due to tunneling effects). We found that, 
 at large distances, the behavior 
 is the same as in Hawking-Page scheme. This is due to the fact that the 
 cosmological constant dominates in the deep infra-red. But, there is a 
 crucial difference in the ultra-violet regime which is controlled by the 
 minimal length $\sqrt{\theta}$. First of all, regular BHs are multi-horizon 
 structures, contrary to usual Schwarzschild BH, and thus are endowed with
 extremal  configurations. 
 The existence of degenerate horizon makes the thermodynamical situation 
 different: as soon as the temperature of the thermal bath is $T> 0$ the 
  near-extremal BHs of minimal mass $M> M_0$ are present.
 As a consequence, there is no pure radiation phase, rather there are a 
 \textit{plethora} of different BH states ranging from single, near-extremal 
 configuration to multiple BH states with the same temperature, but different 
 stability conditions.\\
 Another very interesting feature of regular BHs is that in the strong-coupling
 regime there is always only a single BH at any temperature.\\
 
 While we were completing this paper, we came across of \cite{Nicolini:2011dp}
 where the same model has been studied in $4D$. The ideas presented in that paper follow
the  same reasoning as in \cite{Chamblin:1999tk}. This approach, however, looks more like
 a mathematical analogy based on the similarity between the Van der Walls $P$-$V$
 curve, for a real gas, and the $\beta=1/T_H$ curve as a function of $r_+$. Although
one may accept that this analogy is suggestive, it is hardly an exact
correspondence since $\beta$ is interpreted as ``~pressure~''. Also it mixes  
intensive and extensive thermodynamical variables as noticed in \cite{Kubiznak:2012wp}.\\
Our intention in this paper was not to rely on any formal analogy, but to develop a physically 
based picture of the thermodynamical evolution of regular SAdS BHs studying the temperature
and off-shell free energy. A very important out-come of our analysis is that 
relevant information can be extracted both from the graph of the Hawking temperature and off-shell
free energy. This gives alternative possibility with respect to finite temperature quantum
field theory where information can be recovered only from the effective potential playing
the role of off-shell free energy.

\end{document}